# The Accordiatron: A MIDI Controller For Interactive Music

**Michael Gurevich**
Center for Computer Research in Music and Acoustics
Department of Music, Stanford University
Stanford, California 94305-8180, USA
gurevich@stanford.edu

**Stephan von Muehlen**
Joint Program in Design
Stanford University
P.O. Box 11922
Stanford, California 94309, USA
svm@stanford.edu
http://www.stanford.edu/~svm/accordiatron.html

**ABSTRACT**
The Accordiatron is a new MIDI controller for real-time performance based on the paradigm of a conventional squeeze box or concertina. It translates the gestures of a performer to the standard communication protocol of MIDI, allowing for flexible mappings of performance data to sonic parameters. When used in conjunction with a real-time signal processing environment, the Accordiatron becomes an expressive, versatile musical instrument. A combination of sensory outputs providing both discrete and continuous data gives the subtle expressiveness and control necessary for interactive music.

**INTRODUCTION**
The Accordiatron was initially conceived as a class project for a course in human-computer interfaces at the Center for Computer Research in Music and Acoustics (CCRMA) at Stanford University. Its designers brought their respective expertise in product design and interactive performance together to create a new device that aims to combine aesthetic appeal, comfortable operation and useful output from gestural sensing.

**THE NEED FOR NEW CONTROLLERS**
There exists a highly debated problem in interactive music that is characterized by a lack of correlation between performance gestures and their musical consequences. This can create a difficulty in perception and indeed reception of interactive music; the music can often be reduced to a "challenge" of seeking some relationship between the actions of the performer and the resulting sound. The listener can feel a gap between him or herself and any meaningful purpose to the music. The remedy to this problem is certainly partly the responsibility of the composer, who, if concerned, should try to make more accessible music, but there are some limitations inherent in the instruments used in these compositions that prevent the bridging of this gap.

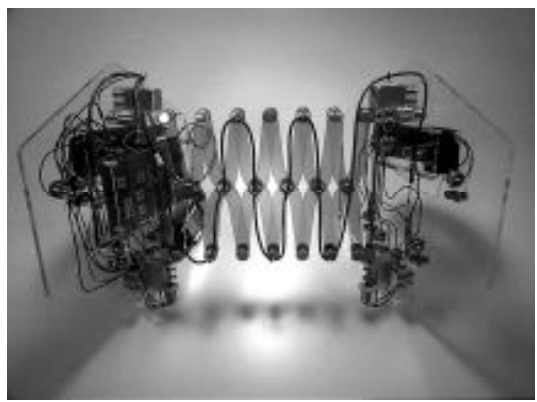

**Figure 1 - The Accordiatron**

The practice of employing traditional instruments in interactive performance situations, either by retrofitting the instrument with various sensors or simply tracking its audio output, can prove to be problematic for the audience. Musical listeners have an ingrained urge to identify the source of a sound. They have a similar expectation of the nature of the sound that will be produced by an instrument. Presently, people have little difficulty accepting the fact that a computer or a recording can be a sound source and have come to expect all sorts of sounds from these media. However, in the case where a traditional instrument is used as a controller for computer music, these expectations are abandoned. The paradox in seeing a familiar instrument and hearing sounds other than those associated with the instrument can be difficult to overcome. In such a case, there is no apparent source for the sound; only this contradiction. The listener is left with the task of trying to associate specific performance gestures with sonic events. In itself, this can be a novel game, but it is certainly not the goal of every composer of interactive music. Traditional instruments are limited in this context in that they are designed for the sounds they produce, not the gestures associated with playing them. There is therefore a need, in the authors' view, for new gestural controllers designed specifically for interactive music situations.

## DESIGN INSPIRATION

Upon setting out to create a new controller that could function as an electronic instrument, the authors devised a set or requirements for the new device. Experience with other gestural controllers for interactive music, such as the Theremin, generated several criteria for the nature of a new controller, while traditional instruments determined others.

### Multi-Dimensionality

Traditional instruments are capable of producing variations in numerous dimensions such as pitch, loudness, timbre, articulation, etc., all of which can be altered at varying rates and applied to determine auditory parameters in real time. In order to be versatile and as useful as an acoustic instrument for interactive applications, an electronic controller similarly needs to provide multi-dimensional data (several unique streams) capable of simultaneously controlling multiple sonic parameters. This has been one limiting factor of the Theremin as a controller, rather than as an instrument: the performer has control in only two dimensions, pitch and loudness.

### Discrete and Continuous Data

A further requirement was that the sensory output provide a combination of discrete and continuous data. This is necessary to suit diverse interactive musical applications, where both singular, isolated events and smooth changes can occur. This is again a limitation of other controllers, like the Theremin, where the only outputs are continuously changing.

### Visual Intrigue

In order to attempt to satisfy a live audience's desire for visual and auditory correspondences as described above, a new controller needed to have strong visual appeal. A performance instrument should be interesting to watch as well as to hear, otherwise part of the purpose of live performance is lost.

### Constrained Motion

The final requirement was that the gestures of the performer be constrained in some way. While free motion in space may be visually expressive, it can be difficult to translate to musically useful information. Furthermore, there are several such systems in existence. The goal in this case was to design an instrument that could be physically manipulated. Motion with limited constraint allows for the control and precision required for musical expression.

## THE SQUEEZE BOX PARADIGM

The paradigm of the squeeze box satisfies many of these design requirements. A number of buttons on the instrument allow for discrete events, while the squeezing motion of the hands and arms is continuous. Sensing the unique button presses as well as the constrained hand motions in several directions allows for multi-dimensional output. The squeeze box was also a compelling starting point because of the expressive physical engagement of the performer and the subsequent value for live interaction. Capturing gesture as an input for electronic music seems to be a very effective means of overcoming the distance between the audience's expectations for a musical performance and the static nature of computer hardware.

In designing the Accordiatron, three types of motion borrowed from the squeeze box were chosen for continuous positional sensing. These are the distance between two ends panels attached to the hands (amount of "squeeze" or "pull"), and the rotation of each hand in relation to the mechanism connecting the two end panels. These are natural motions associated with playing a squeeze box, and while the angle of the hands is not essential in a performing role, it occurs nonetheless. The discrete sensors in the Accordiatron are in the form of a series of buttons, borrowed directly from the design of a concertina.

## THE ACCORDIATRON

The way in which the Accordiatron breaks from the mould of its acoustic counterpart is in the flexibility to apply the sensory data from these motions and buttons to any number of sonic parameters. In a traditional concertina, button presses determine which note the instrument sounds, while the velocity of the squeezing action (and hence the pressure applied by the bellows to the reed) determines the loudness (and incidentally the timbre) of the resulting sound. The output of the Accordiatron could easily produce the same results, simply by mapping button presses to note events in a synthesizer and mapping the rate-of-change of the distance between the two ends to the overall loudness. However, the Accordiatron is capable of much more. In conjunction with a real-time audio processing environment, it can control any aspect of sound.

### The Communication Medium

This is achieved by the use of the MIDI protocol [1]. The Accordiatron outputs MIDI data from a standard 5-pin MIDI connector and can be connected to any device equipped to receive MIDI. It specifically uses control change messages on channels 1-3 for the three continuous controllers, using controller numbers 16-18, which are designated as general purpose controllers. Each is assigned a value from 0 to 127 (the standard 7-bit encoding of MIDI) based on the sensed position on the controller. Only changes in position are transmitted so that filtering of repetitive data by a receiving device is unnecessary. The MIDI standard therefore prevents truly continuous data transmission, but the MIDI transmission rate allows for roughly 1000 total events per second.

Button presses are encoded as note-on messages, each of the 10 buttons with a unique note number from 60-69, on a unique channel from 4-13. A note-on with velocity 127 is transmitted when a button is pressed, and another with velocity 0 upon release. No data after the initial note-on is sent while a button is held down. Button presses are therefore transmitted as truly discrete events.

The Accordiatron presently contains one extra non-performance button that transmits a note-on number 70 on channel 14. This can be used for mode changes or as an additional trigger for setup. Future Accordiatron versions could contain two more such buttons either for transmitting messages on channels 15 and 16, or for changing aspects of the Accordiatron's internal program.

**The Guts**

At the heart of the Accordiatron is the Parallax Basic Stamp II programmable microprocessor [2]. The Accordiatron's program, based on code by Craig Sapp [3], is therefore flexible and can easily be modified in future versions. The continuous sensory input comes from 3 potentiometers that are mechanically rotated by the motions of the performer. The distance between the two ends is sensed by a logarithmic pot attached to one of the linkages with its angle-of-rotation changing as the mechanism opens and closes. The other two potentiometers linearly measure the rotation of each the end panels. These yield output voltages proportional to the motions, whose ranges are adjusted in three operational amplifier circuits, in order to maximize the 0-5V range of the analog-to-digital converters (ADC's). The sensory data is digitized by two 12-bit 8-channel MAXIM ADC's, the first 5 bits of which are ignored to suit the MIDI standard. The button presses are similarly digitized based on high-low voltage changes. The entire system is run from a 9V battery.

**DESIGN**

The starting point for the physical design of the Accordiatron was a traditional squeeze box. It evolved from that model into its final form by incrementally eliminating the non-essential details and resolving the issues that came with the integration of the circuitry and mechanical elements. The approximate dimensions, the kinds of motions the designers wanted to capture, and the combination of continuous and discrete inputs (in the form of buttons) all come from its acoustic counterpart. The reasons for the differences between the two range from prototyping considerations to issues of ergonomics and interface design.

From the squeeze box the designers took the basic layout, two end panels with buttons assigned to each hand and a mechanism between them with some form of continuous control. By isolating the kinds of gestural motion of the performer, a mechanism that could translate that motion to three trim-pots was devised. 1/4 inch acrylic sheet was chosen as the material for the linkages (and end panels) because of its physical and visual characteristics as well as the ease of fabricating the parts with a Computer-Aided Design (CAD) software package and a laser cutter. Ashlar Vellum was the CAD package the designers used in conjunction with Lasercamm's computer-aided machining (CAM) software and laser cutter. Custom hardware was machined from aluminum for the hinges at the end panels and the attachment points of the pots. The remaining hardware consists of off-the-shelf stainless steel and aluminum fasteners used to assemble the various parts.

The end panels each contain nylon hand straps, five buttons, and an offset pad for the palm of the hand. This allows for improved leverage for the fingers and a more natural hand position. The buttons are off-the-shelf parts, but after some user testing they had to be modified with levers hinged over them in order to provide better action and allow for a range of hand sizes.

All of the circuitry is contained on the inside of the right-hand panel. The wiring from the buttons and pot on the left side are carried along the linkages through eye-hooks to the right, where they are soldered along with the rest of the wiring to the circuit board. Stops preventing interference from the mechanism and an additional cover-plate mounted to the end panel protect the circuit board.

The careful consideration of these details in the design serves two purposes. From the standpoint of the user, the Accordiatron feels very much as a musical instrument should. The mechanical movement is smooth, the action of the buttons is very similar to valves or keys, and the feel of the Accordiatron in the hands of the musician is quite natural. For the audience the exposed electronics and mechanism and their movements become a point of interest during a performance and give visual clues linking the viewer's expectations with the sounds being created.

**APPLICATIONS**

The Accordiatron is well suited for interactive performance applications. The following URL contains both video and audio examples of a performance of a piece for the Accordiatron composed and performed by Michael Gurevich using the Max/MSP programming environment:
http://www.stanford.edu/~svm/accordiatron.html

This piece demonstrates some of the capabilities of the Accordiatron. The buttons in this case are mapped to wavetable synthesis using accordion samples, on one hand to single notes, on the other to clusters of notes. These are convolved in real time with a sound file. A cross-fade from the raw sound of the two sources through the convolution of the two is controlled by the distance between the two ends. The angles-of-rotation of the end panels control various other parameters, including the playback rate (pitch) of the accordion samples, reverb room size and several kinds of modulation.

This certainly does not represent the full extent of the musical possibilities of the Accordiatron, but is an indication of its usefulness as a performance instrument in this interactive context.

**ACKNOWLEDGMENTS**

We would like to thank Jonathan Berger, Bill Verplank, Max Mathews, Tamara Smyth and Aaron Hipple for their guidance and assistance in the realization of this project.